\documentclass[fleqn,usenatbib]{mnras}

\usepackage{newtxtext,newtxmath}

\usepackage[T1]{fontenc}
\usepackage{ae,aecompl}

\usepackage{graphicx}	
\usepackage{amsmath}	
\usepackage{amssymb}	

\newcommand{\se}[1]{\S\ref{sec:#1}}

\newcommand{\Fig}[1]{Figure~\ref{fig:#1}}

\newcommand{\be}{\begin{equation}}
\newcommand{\ee}{\end{equation}}
\newcommand{\bea}{\begin{eqnarray}}
\newcommand{\eea}{\end{eqnarray}}

\newcommand{\msun}{{\rm M}_\odot}
\newcommand{\Msun}{M_\odot}

\newcommand{\ifm}[1]{\relax\ifmmode#1\else$\mathsurround=0pt #1$\fi}
\newcommand{\kms}{\ifmmode\,{\rm km}\,{\rm s}^{-1}\else km$\,$s$^{-1}$\fi}
\newcommand{\hmpc}{\,\ifm{h^{-1}}{\rm Mpc}}

\newcommand{\Mpc}{\,{\rm Mpc}}

\newcommand{\ltsima}{$\; \buildrel < \over \sim \;$}
\newcommand{\lsim}{\lower.5ex\hbox{\ltsima}}
\newcommand{\gtsima}{$\; \buildrel > \over \sim \;$}
\newcommand{\gsim}{\lower.5ex\hbox{\gtsima}}

\def\la{\langle}

\def\omm{\Omega_{\rm m}}

\def\omb{\Omega_{\rm b}}

\def\Ms{M_*}

\usepackage{color}

\usepackage{xcolor}
\usepackage{soul}

\title[MZR at DAWN]{Weak Evolution of the Mass-metallicity Relation at Cosmic Dawn in the FirstLight Simulations}

\author[Langan et al.]{Ivanna Langan,$^{1,2}$\thanks{E-mail: langan.annavi@gmail.com}
Daniel Ceverino,$^{1,3,4,5}$
Kristian Finlator $^{1,6}$
\\
$^{1}$Cosmic Dawn Center (DAWN) \\
$^{2}$Sorbonne Universit\'{e}, facult\'{e} des Sciences et Ing\'{e}nierie, 4 place Jussieu, 75005 Paris, France\\
$^{3}$Niels Bohr Institute, University of Copenhagen, Lyngbyvej 2, 2100 Copenhagen $\mbox{\normalfont\O}$, Denmark \\
$^{4}$Universit\"{a}t Heidelberg, Zentrum f\"{u}r Astronomie, Institut f\"{u}r Theoretische Astrophysik, Albert-Ueberle-Str. 2, 69120 Heidelberg, Germany\\
$^{5}$Departamento de F{\'i}sica Te{\'o}rica, Universidad Aut{\'o}noma de Madrid, 28049 Madrid, Spain \\
$^{6}$New Mexico State University, Las Cruces, NM, USA
}

\date{Accepted XXX. Received YYY; in original form ZZZ}

\pubyear{2020}

\begin{document}
\label{firstpage}
\pagerange{\pageref{firstpage}--\pageref{lastpage}}
\maketitle

\begin{abstract}
Little is known about the mass-metallicity relation (MZR) in galaxies at cosmic dawn. Studying the first appearance of the MZR is one of the keys to understanding the formation and evolution of the first galaxies. In order to lay the groundwork for upcoming observational campaigns, we analyze 290 galaxies in halos spanning $M_{\rm h}=10^9$--$10^{11}\msun$ selected from the FirstLight (FL) cosmological zoom simulations to predict the MZR at $z=5$--8. Over this interval, the metallicity of FL galaxies with stellar mass $M_*=10^8\msun$ declines by $\leq0.2$ dex. This contrasts with the observed tendency for metallicities to increase at lower redshifts, and reflects weakly-evolving or even increasing gas fractions. We assess the use of the $R3$ strong-line diagnostic as a metallicity indicator, finding that it is informative for $12+\log(O/H)<8$ but saturates to $R3\approx3$ at higher metallicities owing to a cancellation between enrichment and spectral softening.
Nonetheless, campaigns with \emph{JWST} should be able to detect a clear trend between
$R3$ and stellar mass for $M_*>10^{7.5}\msun$. We caution that, at fixed metallicity,
galaxies with higher specific star formation show higher $R3$ owing to their more intense
radiation fields, indicating a potential for selection biases.

\end{abstract}

\begin{keywords}
galaxies: evolution -- galaxies: formation  -- galaxies: high-redshift 
\end{keywords}



\section{Introduction}

Little is known about the abundances of elements heavier than helium in the primeval galaxies of the early Universe.
These elements are thought to be produced in the first supernovae explosions~\citep{BrommYoshida11}.
Therefore, the generation of heavy elements was quickly linked to star-formation processes and galaxy growth, driven by gas inflows, outflows and merging.
In fact, the evolution of the metallicity, the content of heavy elements relative to hydrogen and helium, gives strong insights about these processes \citep{Maiolino2019}.

The galaxy average gas-phase metallicity shows a strong scaling relation with the galaxy stellar mass. The mass-metallicity relation (MZR) has been observed from the local Universe ($z=0$) to cosmic noon ($z\simeq3.5$; see \cite{Maiolino2019} for a review of observational efforts). 
There is a consensus that the MZR evolves at $z<3.5$ in the sense that, at fixed stellar mass, the metallicity declines with redshift.
Does this trend continue to even higher redshifts? There are no observational estimates of metallicities at cosmic dawn ($z\geq5$) with the exception of GRB afterglows \citep{Berger07}. 
Measurements at lower redshifts ($z\leq3$) are mainly based on strong emission lines, such as [OIII]$\lambda$5007. At cosmic dawn, these optical lines are redshifted out of the spectral range of current spectrographs. 
The James Webb Space Telescope (\emph{JWST}) will soon open a new window into this range, allowing the MZR to traced to much earlier times.
Therefore, it is now the best time to make theoretical predictions about the evolution of the MZR at cosmic dawn.

Cosmological simulations have become very advanced tools to study the evolution of the MZR \citep{Finlator08, DeRossi17, Torrey18, Dave19}. However, there are only a few cosmological simulations that provide results at cosmic dawn \citep{PaperI, Barrow17, Rosdahl18, Finlator18, Torrey2019, Pallottini19, Katz19, Ma19}. Most of these works achieve a high resolution using the ``zoom-in" technique, which concentrates all computational power on a few select halos.

The main advantage of zoom simulations over full-box simulations is that they treat the baryon cycle with much more realism. For example, they model both the emergence of star formation-driven outflows and the small-scale interactions between outflows, the host galaxy's interstellar medium, and the circumgalactic medium more accurately due to their higher resolution and more detailed models of star formation and feedback. As these processes of cooling, star formation, ejection, mixing, and re-accretion are the processes that govern the baryon cycle and the metal enrichment within galaxies, it is important to compare predictions emerging from both simulation techniques.

The traditional drawback to zoom-in simulations is that they do not reveal the ensemble statistical properties of galaxy populations such as the MZR and its scatter. Consequently, most of works focus on the formation of few galaxies.
The FirstLight  \citep[Paper I]{PaperI} database of $\sim$300 zoom-in simulations overcomes this limitation. Moreover, FirstLight galaxies agree well with a variety of observations at $z=5-8$, such as the UV luminosity function or the stellar mass function (Paper I), the SFR-$M_*$ relation \citep[Paper II]{FirstlightII} and $M_*$-Luminosity relations \citep[Paper III]{FirstlightIII}. Thus, the FirstLight database is the ideal laboratory to study the predicted MZR at $z\geq5$.

This paper has two well-defined goals.
After the description of simulations (Section \se{runs}), we characterize the MZR at different redshifts at cosmic dawn (Section \se{MZR}) and study its evolution (Section \se{evo}). Then, we provide metal-sensitive observables (Section \se{R3}) that can be used in the calibration of future measurements of metallicity at cosmic dawn. Section \se{discussion} ends with the summary and discussion.

\section{Simulations}
\label{sec:runs}

This paper uses a complete mass-selected subsample from the FirstLight database of simulated galaxies, described fully in Paper I.
The subsample consists of 290 halos with a maximum circular velocity ($V_\mathrm{max}$) between 50 and 250 $\kms$, selected at $z=5$.
The halos cover a mass range between a few times $10^9$ and $10^{11} \ \msun$.
This range excludes more massive and rare halos with number densities lower than $\sim 3 \times 10^{-4} (h^{-1} \Mpc)^{-3}$, 
as well as small halos in which galaxy formation is extremely inefficient.
 
The target halos are initially selected using low-resolution N-body only simulations of two cosmological boxes with sizes 10 and 20 $\hmpc$, assuming WMAP5 cosmology with $\omm=0.27$, $\omb=0.045$, $h=0.7$, $\sigma_8=0.82$ \citep{Komatsu09}.  
We select all distinct halos with $V_\mathrm{max}$ at $z=5$ greater than a specified threshold, log $V_{\rm cut}=1.7$ in the 10 $\hmpc$ box and log $V_{\rm cut}=2.0$ in the 20 $\hmpc$ box.
Initial conditions for the selected halos with much higher resolution are then generated using a standard zoom-in technique  \citep{Klypin11}.
 The DM particle mass resolution is $m_{\mathrm{DM}} = 10^4 \ \msun$. The minimum mass of star particles is $100 \ \msun$.
 The maximum spatial resolution is always between 8.7 and 17 proper pc (a comoving resolution of 109 pc after $z=11$).
  
 The simulations are performed with the  \textsc{ART} code
\citep[][Paper I]{Kravtsov97,Kravtsov03, Ceverino09, Ceverino14}, which accurately follows the evolution of a
gravitating $N$-body system and Eulerian gas dynamics using an adaptive mesh refinement (AMR) approach.
Besides gravity and hydrodynamics, the code incorporates 
many of the astrophysical processes relevant for galaxy formation.  
These processes, represented via subgrid 
physical prescriptions, include gas cooling due to atomic hydrogen and helium, metal and molecular hydrogen cooling, photoionization heating by a constant cosmological UV background with partial 
self-shielding, star formation and feedback (thermal+kinetic+radiative), as described in paper I.
The simulations track metals released from SNe-Ia and from SNe-II, using supernovae yields that approximate the  results from \cite{WoosleyWeaver95}. 
These values are given for gas cells and star particles as described in \cite{Kravtsov03}.  

We assume that the unresolved nebular region around each star particle shares the same mass ratio of metals produced in SNII explosions as in the star particle (Paper III). The galaxy metallicity is defined as the mass-weighted average nebular metallicity including all star particles younger than 100 Myr.
Using the supernovae yields included in the simulation (\cite{WoosleyWeaver95}), our definition of solar metallicity corresponds to $Z_\odot$ = 0.02 and ${\rm log} (O/H) + 12 = 8.9$. This normalization differs from other simulation works, such as \citet{Torrey2019}, which assume a slightly lower value of 8.6. This systematic difference in normalization is comparable to the $\sim$factor-of-two uncertainties in observational metallicity calibrations~\citep{Kewley08}. It does not impact our current study, which focuses on the relative evolution of the MZR at cosmic dawn.

The luminosities of metal-sensitive emission lines are extracted from the publicly available SEDs described in Paper III. In summary, the SEDs of the simulated galaxies are computed using publicly available tables from the Binary Population and Spectral Synthesis (BPASS) model \citep{Eldridge17} including nebular emission \citep{XiaoStanway18}. 
We combine all individual SEDs coming from all star particles within each galaxy.
The nebular emission originates in regions around star particles younger than 100 Myr.

 \begin{figure*}
	\includegraphics[width=2.1 \columnwidth]{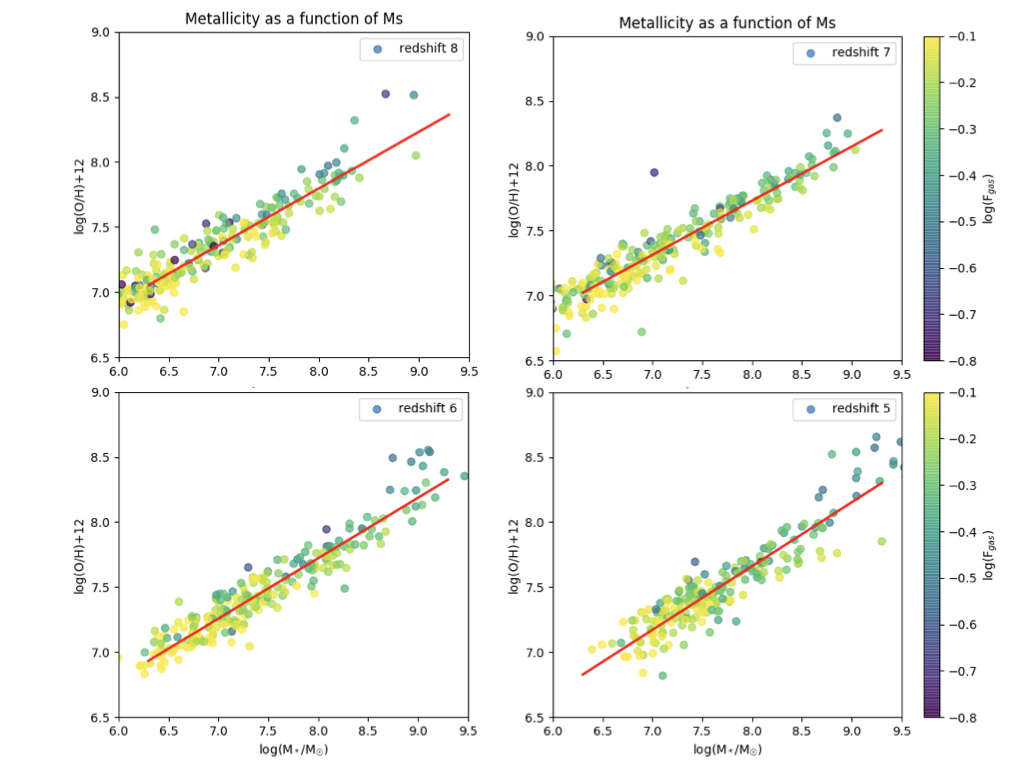}
		\caption{Metallicity as a function of the stellar mass from redshift $z=5$ to 8. The color bar represents the gas fraction. The red lines mark the linear fit. There is weak evolution of this scaling relation at cosmic dawn.}
	  \label{fig:FMR}
\end{figure*}

As described in \cite{XiaoStanway18}, we compute the line luminosities of a single nebular region around a star particle by assuming a constant nebular density of $n_{\rm H}=300 \ {\rm cm}^{-3}$, because the simulations do not resolve the nebular regions around young stars where most of the nebular light is emitted. This is the value normally used in the literature \citep{Steidel16}. Denser H II regions,  $n_{\rm H}=1000 \ {\rm cm}^{-3}$, give similar results for the SSP metallicities considered in this paper \citep{XiaoStanway18}. Next, we measure the ionization parameter at the Str\"{o}mgren radius. The Str\"{o}mgren radius for each star particle is calculated using the assumed nebular density and the properties of the stellar population. Finally, we use a publicly available grid of Cloudy models to derive the luminosities of the most prominent optical and UV emission lines. More details can be found in Xiao et al. 2018.

Radiative transfer effects and dust attenuation are not included in the analysis. These could affect the rest-frame optical line luminosities of the most massive galaxies, $\Ms > 10^8 \Msun$. Paper III has shown that the computed dust-free UV slope ($\beta$) is consistent with observations \citep{Bouwens14} for galaxies fainter than $M_{\rm UV}\simeq -19$. In other words, available observations are consistent with the assumption that dust obscuration has a minor impact on the luminosity ratios of rest-frame optical lines for most of the FirstLight sample.

\section{Results}

\subsection{The Mass-Metallicity Relation}
\label{sec:MZR}

In Fig.~\ref{fig:FMR}, we show that a tight MZR is predicted to exist throughout $z=5-8$. Moreover, it evolves very little with redshift. We do not see a change in the slope, as reported by \citet{Torrey2019} at $\Ms\simeq10^9 \Msun$. The shallower slope in FirstLight results from the minimum wind velocity imposed in the recent Illustris model \citep{Pillepich18b}. This assumption reduces the mass loading factor (that is, the gas outflow rate in units of SFR)
in comparison to what occurs without such a velocity floor \citep{Torrey2019}. 
A reduced loading factor naturally increases the galaxy metallicity at a fixed stellar mass because more metals are retained in the ISM rather than expelled \citep{Lilly13,Finlator17}.

Observations indicate that the scatter in the MZR correlates with other galaxy properties including gas fraction and SFR \citep{Maiolino2019}.
Previous cosmological full-box simulations have reported this correlation at lower redshifts, $z\leq4$ \citep{DeRossi17, Torrey2019, Dave19}.
In order to explore whether this is expected even at $z\geq6$, we use colors to indicate the gas fraction $F_{\rm gas}$, defined as $F_{\rm gas} \equiv M_\mathrm{\rm gas}/(M_\mathrm{\rm gas}+M_\mathrm{*}$) in Fig.~\ref{fig:FMR}. Indeed, it does seem that a residual correlation between $Z$ and $F_{\rm gas}$ contributes significantly to the MZR scatter.
For example, galaxies with higher-than-average gas fractions, F$_{\rm gas}\simeq 0.7$, have preferentially lower metallicities, at a fixed stellar mass of $10^8 \Msun$. In the same mass bin, galaxies with lower-than-average gas fractions, F$_{\rm gas}\simeq 0.3$, have metallicities higher by 0.2 dex than average.
As shown in Paper II, a higher gas fraction correlates with a higher star formation rate for a given mass and redshift. Therefore, the FirstLight simulations predict that the fundamental mass-metallicity relation \citep{Ellison2008, Maiolino2019} has already emerged by cosmic dawn. 



\subsection{Weak evolution of the MZR}
\label{sec:evo}

\begin{figure}
    \begin{center}
        \includegraphics[width=\columnwidth]{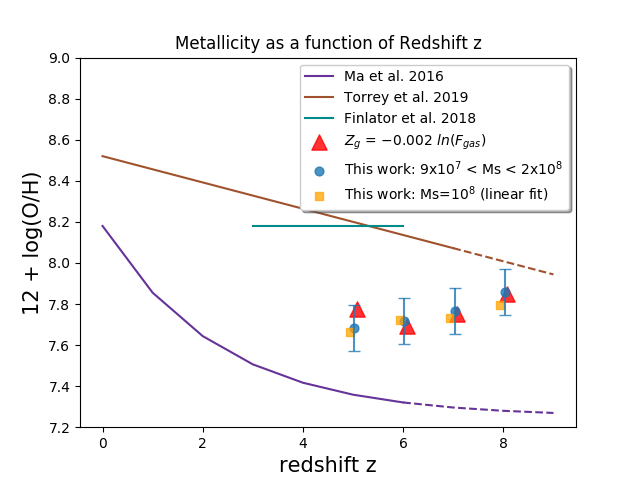}
        \caption{The evolution of metallicity at fixed stellar mass, $M_* = 10^8 \Msun$. Blue circles and yellow squares indicate FL predictions while colored curves indicate predictions from other simulations (dashed lines indicate an extrapolation in redshift). Red triangles show the predicted evolution if the effective yield is fixed to $y_{\rm eff}=0.002$ and $F_{\rm gas}(z, M_*=10^8\Msun)$ is adopted directly from our simulations (section \se{evo}). Note that yellow squares and red triangles are offset slightly in redshift for clarity.}
        \label{fig:comparison}
    \end{center}
\end{figure}

\Fig{comparison} (blue circles) shows the average metallicity and its scatter (vertical error bars) within a narrow mass bin centered at $\Ms=10^8 \Msun$ at different redshifts across cosmic dawn. These average values agree well with the linear fits used in \Fig{FMR} (yellow diamonds).
They show weak evolution of the MZR at these redshifts.
In fact, there is even a hint that metallicity declines by $\sim$0.15 dex from $z=8$ to 5, although the decline is similar to the intrinsic scatter of the relation.

We can show that this weak evolution is driven by weak evolution in $F_{\rm gas}$ by invoking the \emph{effective yield} \citep{Garnett02, Dalcanton07}. If a galaxy is a closed-box system, then the gas metallicity
    \begin{equation}
        Z_g \equiv M_{\rm metals}/M_{\rm gas}
    \end{equation} 
is a function of $F_{\rm gas}$ and the intrinsic stellar yield, $y=0.02$, which expresses the ratio of the mass of new metals released into the ISM to the mass in long-lived stars. In the more general case where inflows and outflows are permitted, $y_{\rm eff}$ quantifies their net impact on a galaxy's chemical content:
    \begin{equation}
        Z_g=-y_{\rm eff} \ {\rm ln}(F_{\rm gas})
    \label{eq:eff}
    \end{equation} 
Under the assumption that the net impact of inflows and outflows is constant throughout our redshift range and stellar mass bin, $y_{\rm eff}$ is constant and the MZR's evolution in time at a fixed stellar mass is driven by evolution in $F_{\rm gas}$.
The average gas fractions increase from $F_{\rm gas}\simeq 0.45$ at $z=8$ to $F_{\rm gas} \simeq 0.55$ at $z=6$. In order to verify that this suffices to drive the predicted metallicity evolution, we use red triangles in \Fig{FMR} to plot the quantity $-0.002\ln(F_{\rm gas})$. This model closely tracks the FirstLight simulations throughout $z=8\rightarrow6$, supporting the idea that slowly-increasing gas fractions indeed dominate early metallicity evolution within the full simulation. From $z=6\rightarrow5$, the agreement is weaker, although the constant effective yield model still agrees with the simulation to within the simulated scatter. For context, we note that the assumed value $y_{\rm eff}=0.002$ is similar to the values found in local galaxies of similar mass \citep{Tremonti04}. 
The effective yield is 10 times smaller than the intrinsic yield, indicating that the interstellar media of reionization-epoch galaxies were highly dynamical environments characterized by a vigorous interplay between metal-rich outflows and pristine inflows.

The slowly-increasing gas fractions that arise in our simulations predict that, prior to $z\simeq5$, galaxies build up their gas reservoirs because the rate at which star formation and outflows process their ISM lags the gas accretion rate. In this regime, dilution from pristine inflows dominates over enrichment, leading to constant or even slowly-decreasing gas-phase metallicities \citep[see also][]{Wu19}.

As the MZR at $z>4$ remains unconstrained observationally, we focus on a comparison with predictions from other theoretical models. \cite{Ma16} show results up to $z=6$ and \cite{Torrey2019} extend it to $z=7$. In order to facilitate comparison with FirstLight predictions, we extrapolate their fitting relations to our range of redshifts. In doing so, we normalise their predictions to the same solar abundance that we assume (See \se{runs}).
Our metallicity values lie in between these predictions. The values are close, within $\sim$0.5 dex. This is encouraging because these simulations use slightly different definitions of metallicity but they claim to reproduce the same observed values at lower redshifts, $z\leq2$. 

However, there are some important differences in the evolution of the MZR with respect to these previous works.
In \citet{Torrey2019}, we can see that the metallicity strongly decreases with redshift at all times. 
This evolution is driven by the artificial velocity floor imposed on the outflows, as described above. 
Indeed, their massive galaxies, $\Ms \geq10^{9.5} \Msun$, start to show a flattening in their metallicity evolution with redshift at $z\simeq5$. This is because more massive galaxies have outflows with higher velocities and they are not affected by this velocity floor. 

The simulations by \cite{Finlator18} show no metallicity evolution from  z=3 to 6, roughly consistent with our findings at higher redshifts. Although their wind model is similar to \cite{Torrey2019}, it does not include the velocity floor. Most probably, this is why their metallicity does not decrease with redshift as in previous works. This highlights the importance of galactic outflows in the evolution of the metal content in galaxies.

The simulations by \citet{Ma2016} predict a flattening of the metallicity evolution with increasing redshift, roughly consistent with the weak evolution in FirstLight. 
However, the predicted metallicities are systematically lower by 0.3 dex. It is likely that the outflows in these simulations are far more efficient in removing metals from small galaxies. Indeed,~\citet{Agertz19} have shown that this model struggles to reproduce the observed plateau of stellar abundances in the faintest dwarfs in the local volume. 

These comparisons highlight the importance of metallicity as a sensitive test of feedback models. Two questions motivated by this discussion include. First, when are galaxies predicted to graduate from building up their gas reservoirs to exhausting them? The epoch at which $F_{\rm gas}$ peaks may manifest observationally via a minimum in the MZR normalization. Secondly, how can upcoming measurements of $F_{\rm gas}$ and Z$_{\rm gas}$ be used to identify this transition? Over the next decade, observations of galaxies and their circumgalactic media at cosmic dawn will distinguish between these models, illuminating how the MZR first emerged.



\subsection{Metallicity Indicators}
\label{sec:R3}

We now assess the strong rest-frame optical emission line diagnostics that will soon be used to test the predictions in the previous sections. These lines are widely-used to infer the metallicities of low-redshift galaxies because they are bright and easily-accessible using ground-based measurements. At reionization-epoch redshifts, however, only \emph{JWST} can open up rest-frame optical diagnostics of galaxy evolution.

From the published mock optical spectra of the galaxies in the FirstLight database (paper III), we measure the luminosities of [OIII]$\lambda$5007, OII ($\equiv$[OII]$\lambda$3727+[OII]$\lambda$3729), and H$\beta$. From these fluxes, we compute the metallicity-sensitive \textit{R2} and \textit{R3} indices following~\citet{Maiolino2019}. As oxygen is the most abundant heavy element, we can compare them with the intrinsic gas-phase metallicities discussed in the previous section. In practice, we have found that the contribution of \textit{R2} to \textit{R23} (which is the sum or \textit{R2} and \textit{R3}) is very small. We therefore focus on \textit{R3}, with the understanding that qualitative results apply also to \textit{R23}.

\begin{figure}
    \begin{center}
        \includegraphics[width=\columnwidth]{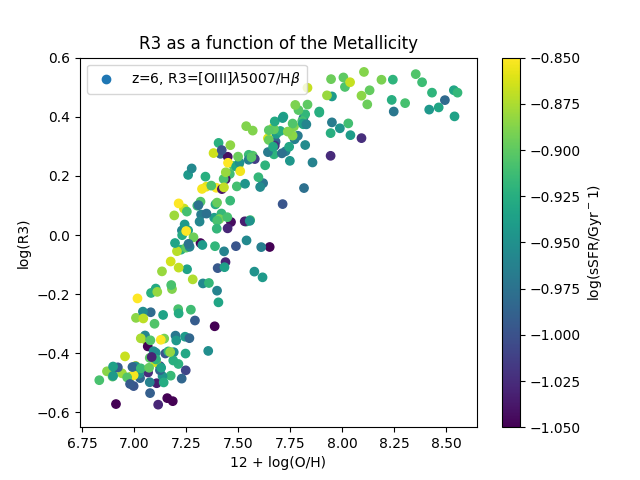}
        \caption{\textit{R3} as a function of the metallicity at redshift $z=6$. The color bar represents sSFR. 
        After a maximum value at around $Z=0.1 Z_\odot$, \textit{R3} strongly decreases at lower metallicities. For a given metallicity bin, \textit{R3} increases with sSFR.}
         \label{fig:R3L}
    \end{center}
\end{figure}

In Fig.~\ref{fig:R3L} we plot \textit{R3} versus the galaxy metallicity.
For $12+\log(O/H)<8$, \textit{R3} increases strongly with gas-phase metallicity. At higher metallicities, it saturates. This behavior reflects a competition between enrichment and ionization. For $12+\log(O/H)<8$, a slight increase in metallicity boosts the oxygen abundance more significantly than it softens the radiation field, yielding an overall increase to \textit{R3}. At higher metallicities, the two effects largely cancel. As a result, \textit{R3} is a useful probe of metallicity only at low metallicities.

The scatter of this relation is mostly driven by the sSFR (sSFR$\equiv {\rm SFR}/\Ms$), which correlates with the ionization parameter (Paper III).
At a given metallicity, galaxies with higher sSFR have systematically higher \textit{R3} values.
These galaxies are usually brighter and easier to observe. This could introduce a bias towards high \textit{R3} values in the observations of the faintest galaxies.

\begin{figure}
    \begin{center}
        \includegraphics[width=\columnwidth]{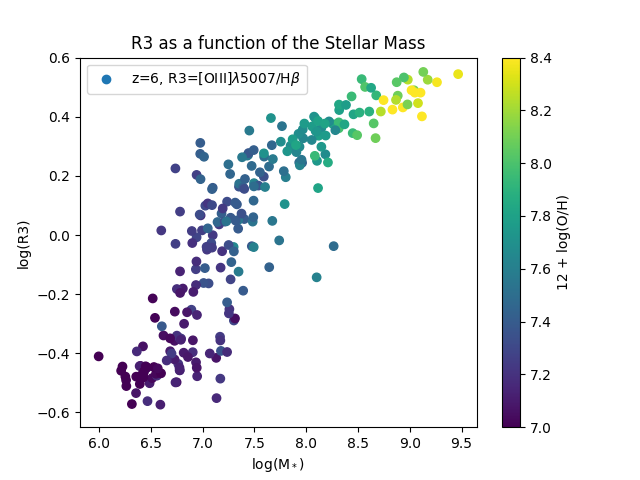}
        \caption{\textit{R3} as a function of the stellar mass at redshift $z=6$. The color bar represents the metallicity in oxygen units. There is drop in \textit{R3} which corresponds to a metallicity around 7.5 and a stellar mass around $10^{7.5} M_\odot$.}
         \label{fig:R3Ms}
    \end{center}
\end{figure}

 In \Fig{R3Ms}, we plot \textit{R3} versus the galaxy stellar mass.
 Here we see a clear trend of increasing \textit{R3} with mass.
 The trend is mostly driven by the galaxy metallicity.
 There is a drop in \textit{R3} at $\Ms \simeq10^7 \Msun$, driven by the low oxygen abundance, as in \Fig{R3L}. 
JWST will only just be capable of observing such low-mass galaxies in blank fields. Ultra-deep fields with a limiting magnitude of $m=31$ in the rest-frame UV would be mass-completed for $\Ms \geq 10^{7.5} \Msun$ at $z=6$ (Paper III). JWST samples of smaller galaxies will be biased towards higher sSFRs, which could prevent detection of the predicted drop in \textit{R3} toward lower masses.

\section{Discussion and Summary}
\label{sec:discussion}

We have used the FirstLight database of cosmological zoom-in simulations to study the mass-metallicity relation at cosmic dawn.
The main results can be summarised as follows:
\begin{itemize}
\item
The MZR, well-known  at low redshifts, is predicted to exist at high redshifts ($z=5$--8).
\item
The scatter of the MZR is driven by the gas fraction.
\item
The average metallicity at a fixed stellar mass evolves weakly---and may even decrease---during the interval $z=8\rightarrow5$. This evolution is driven by slowly-increasing gas fractions during the epoch of gas reservoir buildup.
\item
The emission lines ratios \textit{R3} and \textit{R23} correlate relatively well with both metallicity and stellar mass. However, a secondary dependence on sSFR could introduce observational biases, particularly in samples that are not mass-selected.
\item
There is an abrupt decrease of \textit{R3} at low metallicities and low masses driven by low metal abundances.
\end{itemize}

Galaxy spectra from JWST will open a new window to the rest-frame optical at cosmic dawn \citep{AlvarezMarquez19}. 
New measurements of emission lines ratios at these high redshifts will give us the first direct determination of galaxy metallicities in the early Universe.
However, a lot of work on calibration of these metal tracers will be needed.
This work provides the first attempt to look at these metallicity tracers from cosmological simulations.
Still, systematic uncertainties remain.
The choice of different line-emission models could affect the calibration. The choice of stellar evolution model (in our case, BPASS) is another source of uncertainty. 
Future works will address the systematics associated with these decisions and the effect in the overall normalisation.
However, the reported evolutionary trends are robust to choices that only affect the MZR's normalisation.

Other caveats are intrinsically related to the FirstLight simulations. 
First, they do not follow individual elements, like oxygen, carbon or nitrogen.
Therefore, we rely on published supernovae yields to estimate the oxygen abundances from the total amount of metals produced in core-collapsed supernovae.
This is a good assumption for oxygen because it is mostly generated in these supernovae.
However, other important elements, like carbon or nitrogen, have also secondary production channels, like AGB winds, which are not included.
In addition, radiative transfer effects are only considered at post-processing on unresolved scales. 
The structure and dynamics of nebular regions on very small scales may also affect OIII luminosities \citep{Pellegrini19}.
Dust is also not included, although its effect on \textit{R3}  is small.
Finally, bigger cosmological volumes will be needed to address the metal content in more massive galaxies and/or higher redshifts.

There is still plenty of exciting work to do, things to try and to understand, with respect to the mechanisms of metal production in the early Universe, or more precisely in primeval galaxies. Future works include following the metallicity content of galaxies in the same mass range at $z=5-8$ by computing the 3D metal distributions in time intervals of 10 Myr. As a second step, we will follow the evolution of $M_{\rm metals}$ and $M_{\rm gas}$ for different galaxies at similar masses but different redshifts. We will check if their evolution could explain the mild increase of metallicity we get with redshift. These follow-up works will get new insights about the origin of the elements in the early Universe.


\section*{Acknowledgements}
We thank Xiangcheng Ma for fruitful discussions.
The Cosmic Dawn Center is funded by the Danish National Research Foundation.
The FirstLight simulations were performed at Leibniz Supercomputing Centre and bwHPC.
DC is a Ramon y Cajal researcher.




\bibliographystyle{mnras}
\bibliography{biblio.bib}




\bsp	
\label{lastpage}
\end{document}
